\documentclass{article}

\usepackage[square,sort,comma,numbers]{natbib}
\usepackage[final]{nips_2017}

\usepackage{times}
\usepackage{url}
\usepackage{pbox}
\usepackage{array}
\usepackage{adjustbox}
\usepackage{booktabs}
\usepackage{helvet}
\usepackage{courier}
\usepackage{color}
\usepackage{amsmath}
\usepackage{amssymb}
\usepackage{array}
\usepackage{amsfonts}

\usepackage[utf8]{inputenc} % allow utf-8 input
\usepackage[T1]{fontenc}    % use 8-bit T1 fonts
\usepackage{hyperref}       % hyperlinks
\usepackage{url}            % simple URL typesetting
\usepackage{booktabs}       % professional-quality tables
\usepackage{amsfonts}       % blackboard math symbols
\usepackage{nicefrac}       % compact symbols for 1/2, etc.
\usepackage{microtype}      % microtypography

\title{Detection and Characterization of Illegal Marketing and Promotion of Prescription Drugs on Twitter}
\author{
		Janani Kalyanam\\
		Global Health Policy Institute\\
		University of California, San Diego\\
		\texttt{jkalyana@ucsd.edu} \\
		\And
		Timothy Mackey \\
		Global Health Policy Institute\\
		School of Medicine, Department of Anasthesiology\\
		University of California, San Diego\\
        \texttt{tmackey@ucsd.edu}
}

\begin{document}
\maketitle
\begin{abstract}
		%Prescription drug abuse (PDA) is one of the fastest growing drug-related problems in the United States.  
		%Prior studies have established that one of the major venues for illicit access to prescription 
		%drugs is through online pharmacies.  
		Illicit online pharmacies allow the purchase of prescription drugs online without
		a prescription.
		Such pharmacies leverage social media platforms such as Twitter as a promotion and marketing tool with
		the intent of reaching out to a 
		larger, potentially younger demographics of the population.
		Given the serious negative health effects that arise from abusing such drugs,
		it is important to identify the relevant content on social media and exterminate their presence
		as quickly as possible.  In response,
		we collected all the tweets that contained the names of certain preselected controlled substances
		over a period of 5 months.  We found that an unsupervised topic modeling based methodology is
		able to identify tweets that promote and market controlled substances with high precision.  We also study 
		the meta-data characteristics of such tweets and the users who post them and find that they have 
		several distinguishing characteristics that sets them apart.  We were able to train supervised methods
		and achieve high performance in detecting such content and the users who post them.
\end{abstract}

\section{Introduction}
Prescription drug abuse is the use of a prescription medication 
in a way not intended by the prescribing healthcare professional.  
Prescription drug abuse is a national epidemic, and is the cause 
for the largest percentage of deaths from drug overdose.  
Prior studies have established that illicit 
online pharmacies (IOPs) represent an understudied venue for illegal access of
prescription drugs via the internet\cite{liang2009searching}.  IOPs often provide access to 
controlled substances without a valid prescription. 
Recent studies have also shown that 
IOPs utilize social networking platforms such 
as Twitter to promote and market their products \cite{katsuki2015establishing,katsuki2016digital}.  
Hence, the already challenging
issue of addressing and minimizing the prevelance of prescription drug abuse 
is exacerbated due to the utilization of social media by IOPs \cite{mackey2013digital}.
This is because social media provides 
IOPs a gateway to directly 
market their product to the masses.   While progress in federal 
and state-based efforts have begun to address traditional 
forms of drug diversion linked to non-medical use of prescription drugs (such as through 
the implementation of state Prescription Drug 
Monitoring Programs), information on emerging digital 
environments that enable the non-medical use of prescription 
drug access and abuse behavior, such as through IOPs
that use social media are inadequate.  

As part of an effort to better understand the 
national epidemic of prescription drug abuse and all its associated risk factors, it is important 
to understand the ecosystem of IOPs and how
they utilize social media in the illegal promotion and marketing of 
prescription drugs; specifically identify the relevant content on social
media and the external URLs it may point to, identify the characteristics of such
content and the users who post it etc 
to inform the design and development of methodologies that can identify such marketing content 
as soon as they emerge to prevent further exposure to this
public safety and patient safety hazzard.  The eventual goal of such a study is to build
realtime surveillence systems to detect
rogue content about promoting IOPs as it constantly emerges and evolves on popular social media platforms.

In order to achieve these goals, we mine all the messages containing a predefined list
of prescription drugs from the
popular microblogging service Twitter.  We found that a topic modeling based
methodology could efficiently narrow down on a subset of tweets that promote
IOPs.  Furthermore, we study the similarities and differences between the \emph{rogue}
tweets (and users) and regular tweets (and users) both quantitatively and qualitatively
and find that they differ remarkably in many ways.
This is substantiated through a series of statistical significance tests, time-series analysis,
and by training machine classifiers
whose performance yield high scores on several metrics (accuracy, precision, recall, f1-score).

\section{Data Collection}
\label{sec:data_collection}
We collected messages published on the popular microblogging platform Twitter over a period
of approximately 5 months; from June to November 2015.  Twitter provides a public API that enables
the collection of messages posted by its users via its online platform \cite{TwitterStreamingAPI}.  We used a data collection
methodology involving cloud-based computing services offered by Amazon Web Servies (AWS) and virtual
computers via Amazon \texttt{EC2 t2.micro} instances set to filter and collect tweet objects containing
specific keywords of controlled substances.  

Keywords included the brandnames and international non-proprietary (e.g. generic) names (INN names)
of commonly abused prescription analgesic opioid drugs.  The
INN names of prescription opioid drugs used in this study includes Percocet, Codeine, Oxycodone,
OxyContin, Hydrocodone, Vicodin and Fentanyl.  These keywords were used in conjunction with the Twitter
Streaming API in order to track tweets that contained these keywords.  In order to ensure the
collection of the full volume of data containing these keywords, care was taken so that
the limit rates imposed by Twitter were not reached\footnote{A detailed description of the data collection
methodology has been published in a previous study, and will be cited for camera ready.}.  This generated a total of
620,477 tweets that were collected between the period of June and November 2015.  

\section{Detecting Illegal Marketing on Twitter}
\begin{table*}
		{\small
				\centering
				\begin{tabular}{|c|c|c|c||c|c|c|c|}
						\hline 
						& high      & weed   & \textbf{best}     &           &laughter   & dose      & \textbf{offer}   \\
	  & man       &  ask   & \textbf{free}     &           &medicine   & strippers & \textbf{online}  \\
						codeine & deaths    & slow   & \textbf{pills}    & percocet  &laughing   & lean      & \textbf{super}   \\
			  & australia & dreams & \textbf{online}   &           &agree      & cup       & \textbf{best}    \\
	 & market    & sex    & \textbf{buy}     &           &best       & poured    & \textbf{price}    \\
						\hline
						& \textbf{order} & fda     & approval &           & canada  & \textbf{buy}  & wife  \\
	  & \textbf{online} &children& pediatric &           & drug      &\textbf{online}&doctor\\
						oxycontin & \textbf{quantity}&approves& abuse   & oxycodone  & potential &\textbf{free}&airport  \\
				& \textbf{discount} &11     &police   &           & monopoly &\textbf{pills}& prescriptions\\
	& \textbf{prescription}&young& fake   &           & give    &\textbf{cheap}& west\\
						\hline
						& \textbf{online}& liquid&teeth &           & drug     & pain & \textbf{online}\\
	  & \textbf{buy}  &dragon&wisdon &            &accidental& feel  &\textbf{ pharmacy}\\
						vicodin & \textbf{high}  &flavored&pulled& hydrocodone&died     & good  &\textbf{buy}\\
			  & \textbf{quality} &unicorns&pain&           &toxicity  & makes  & \textbf{order}\\
	 & \textbf{price} &delicious &surgery&           &tyler    & sleep  & \textbf{canada}\\
						\hline
				\end{tabular}
				\caption{This table shows a sample of three example themes from 6 of the drugs (fentanyl 
						did not yield relevant topics). The top five words for each topic are displayed.
						The themes marked in bold were annotated as relevant,
				and the rest were marked irrelevant.}
				\label{table:relevant_irrelevant}
		}
\end{table*}
In order to identify rogue content from our dataset, we begin
by summarizing the content using unsupervised topic models.
Topics models like LDA implicitly learn the word co-occurrence patterns
from the document level word generations \cite{blei2012probabilistic}.  Hence, they suffer immensely
in the presence of sparsity (on an average, tweets are 5 words/document).
We use the Biterm Topic Model instead, where the biterm denotes
an unordered pair of terms in a short context (``apple store", ``C program"),
and the topics are learnt from the biterms aggregated over the entire corpus \cite{cheng2014btm}.
In our experiments, we set the number of topics based on the rule of thumb that 
$k \approx Sparsity(\textbf{X})^{-1}$,
where \textbf{X} is that data matrix of documents-by-terms and manually annotated each of the
topics based on the top-$10$ words from the topic (the number of topics were set to $20$
for the experiments).   
This annotation was carried out in an open ended manner where 
the annotators (a data science expert, and an expert in public health policy)
were asked to mark the themes as relevant based on their judgement of whether
the theme could possibly contain tweets relevant to the marketing and promotions
of IOP.  The themes were to be marked as relevant, irrelevant or in need of
further investigation.  Table \ref{table:relevant_irrelevant} provides summary
of results from this phase.  The inter-annotator
agreement was 1.  This suggests that there are clear indications in the lexical groupings
of the topics that indicate whether or not they are pertinent to the illicit marketing, promotion
and sales of prescription drugs.
In order to further calibrate this methodology, 
once a set of relevant or \emph{rogue} topics were identified,
those tweets whose topic decompositions contained a rogue topic as its most dominant
component were isolated.  This subset of tweets were again manually 
annotated as rogue if the contents of the tweets suggested the marketing and promotion
oF IOPs.  In most cases, more than $90\%$ of the tweets in the \emph{rogue} topics
indeed were annotated to be rogue.
\begin{table*}[h]
		{\tiny
				\centering
				\begin{tabular}{|>{\ttfamily}c||c|c||c|c||c|c||}
						\hline
						\text{DRUG NAMES} & \multicolumn{2}{|c||}{CODEINE} &\multicolumn{2}{|c||}{PERCOCET} & \multicolumn{2}{|c||}{OXYCONTIN}\\
						\hline
						\hline
						feature names& rogue& non-rogue&rogue &non-rogue &rogue &non-rogue \\
						\hline
						retweeted\_status       &0      &0.4131& 0.0571 &0.4688 &0.0036 & 0.2764\\
						retweet\_count          &0.3234 &409.16 & 0.4321&244.13 & 0.0036& 7.2593\\
						favorite\_count         &0      & 386.82& 0& 198.88& 0& 5.1835\\
						in\_reply\_status\_id   &0      & 0.0658& 0& 0.0578& 0& 0.0773\\
						possibly\_sensitive     &0      & 0.102& 0& 0.023&  0& 0.0158\\
						entities\_urls          &1      & 0.1724& 1& 0.1764& 0.9927& 0.7331\\
						entities\_symbols       &0      & 0.0009& 0& 0.0048& 0& 0.0007\\
						entities\_hashtags      &0      & 0.1647& 0& 0.1800& 0& 0.3242\\
						user\_verified          &0      & 0.0022& 0& 0.0025& 0& 0.0227\\
						user\_friends\_count    &12.39  & 1123.05& 10.08& 1281.12& 15.03& 1731\\
						user\_follower\_count   &28.39  & 2666.85& 31.025& 3411.784& 31.97& 7669\\
						user\_statuses\_count   &166995  & 38823.55& 155576& 41665& 159218& 41679\\
						user\_favorites\_count  &0      & 5436.99& 0 & 6054& 0.0109& 4700\\
						\hline
						\hline
						DRUG NAMES & \multicolumn{2}{|c||}{OXYCODONE} &\multicolumn{2}{|c||}{HYDROCODONE} & \multicolumn{2}{|c||}{VICODIN}\\
						\hline
						feature names& rogue& non-rogue&rogue &non-rogue &rogue &non-rogue \\
						\hline
						retweeted\_status       &0      &0.3075& 0.0992     &0.1339 &0.0147 &0.2062\\
						retweet\_count          &0 &16.248& 0.2977          &1.6805 & 0.0147&   6.7211 \\
						favorite\_count         &0      & 8.7775& 0         & 2.5229& 0     & 12.0797      \\
						in\_reply\_status\_id   &0      & 0.0703& 0         & 0.1623& 0     & 0.2034      \\
						possibly\_sensitive     &0      & 0.0107& 0         & 0.0112 &  0   &  0.0134     \\
						entities\_urls          &0.9999      & 0.7109& 0.8777& 0.2692& 1&      0.0769     \\
						entities\_symbols       &0      & 0.0021& 0         & 0.0001& 0     &  0.0011     \\
						entities\_hashtags      &0.0145    & 0.4165& 0.0114 & 0.1953& 0.0221& 0.1528      \\
						user\_verified          &0      & 0.0346& 0         & 0.0084& 0     & 0.0045      \\
						user\_friends\_count    &9.828 & 1511.60& 37.4427   & 1028.98& 12.2022&  1342.68   \\
						user\_follower\_count   &25.518  &11600.15& 110.52  & 3326.318&46.48 &3066.871      \\
						user\_statuses\_count   &158638  & 50960.00& 129878 & 38677&160245 & 31104.78       \\
						user\_favorites\_count  &0.2956     & 3939.64& 36.42& 7172.15& 0.3492& 13508.133     \\
						\hline
				\end{tabular}
				\caption{This table lists the mean values of several Twitter based features for the rogue set of tweets
				and the regular set for all the drugs.}
				\label{table:ttest}
		}
\end{table*}

Next, we wish to characterize tweets with content related to
the promotion and marketing of IOPs, and users who engage with IOP related
content.  The goal for such an analysis is to understand the unique markers
of rogue tweets and users so that intelligent systems can be trained and
deployed to detect their presence and emergence in the social media arena.
As part of this effort, we analyze various twitter based features of the
tweets and the users, and quantitatively (through statistical tests) and qualitatively lay out
their similarities and differences.  Then, we train a machine
classifier to learn the differences between the features and test its
performance on unseen test data.
Table \ref{table:ttest} shows the mean value for
each group of data across drugs for all the features \footnote{We omit fentanyl since
very few rogue tweets were identified.}.  We divide the features
into five semantic groupings; each representing a speific aspect of
user behavior or the characteristics of tweets. 

The \emph{User Engagement Features} such as \texttt{favorite\_count} and \texttt{in\_reply\_to\_status\_id} 
(whether the tweet is a reply to an existing tweet).  
are consistently $0$ for the rogue set of tweets across all drugs.
This suggests that rogue tweets do not invite active engagement from
general users.  The \texttt{retweeted\_status} and \texttt{retweet\_count}
indicate whether the tweet under consideration is a retweet, and the number of times
the original tweet has been retweeted (at the time of data collection).
We observe that these two features
are either $0$ or remarkably low for the rogue set of tweets across all
drugs.  
For those drugs for which the \texttt{retweeted\_status} and \texttt{retweet\_count}
were non-zero, we further investigated the cause for the retweets.  We found
that some of the retweets were propagated by the author of the original tweet itself.
Collectively,
all the user engagement features seem to suggest that users 
in general choose not to engage with content
that might seem suspicious.  

The \emph{Tweet Based Features} are those that describe the content of the tweet.
This includes
\texttt{entities\_urls}, \texttt{entities\_hashtags},
\texttt{entities\_symbols} and \texttt{possibly\_sensitive}.
The \texttt{entities\_urls} feature indicates whether or not a tweet
contains a url embedded in it.
From Table \ref{table:ttest}, we observe that this
feature is consistently 1 or very close to 1 across all the drugs.  This suggests that
one of the sole strategies in the marketing and promotion of IOPs is to direct
the users out of the Twitter domain and into the landing page of the IOPs.

The \emph{User Network Features} are those that indicate the size of a user's
network on Twitter like \texttt{user\_followers\_count} and the 
\texttt{user\_friends\_count}).
From Table
\ref{table:ttest}, we observe that both these features have much smaller
values in the rogue set than the non-rogue set.  On an average, the 
\texttt{user\_friends\_count} for the regular set is approximately $107$x 
higher than the rogue set; and the \texttt{user\_followers\_count} is
approximately $188$x higher in the regular set than the non-rogue set.
This suggests that the users propagating and perpetuating the rogue tweets are fairly
isolated
in the network.
This may be explained by the relative short life of many IOPs, 
which are often removed or become inactive due to a number of 
factors including enforcement activities \cite{mackey2016digital}.

The \emph{User Profile Features} numerically describe the profile of a user 
and includes \texttt{user\_statuses\_count},
\texttt{user\_favorites\_count} and \texttt{user\_verified}.  We note
that the \texttt{user\_statuses\_count} (the total number of status messages
published so far by the user) for the rogue set is on an average approximately $5$x as much
as that of the regular set (this number is in the order of 100000s for the
rogue set, and in the order of 10000s for the regular set).
We also analyzed the data of creation of the user accounts.  More
than $70\%$ of the accounts in the rogue set were created in or
after 2014.  
We also note that there are no verified users in the rogue set of tweets because
the \texttt{users\_verified} feature is consistently $0$ across all drugs.
This indicates that the influential nodes on Twitter do not engage with

We now proceed to train a machine classifier to be able to automatically
classify a tweet as being \emph{rogue} or not based on the features
engineering above.  The data is randomly split into $70\%$ training and $30\%$ test 
splits.  The results were repeated and averaged over $10$ different runs.
The results from logistic regression are summarized in Table \ref{table:classification} \cite{hosmer2000introduction}.
We observe good performance consistently across all drugs.  This provides
sufficient evidence to promise success of real world systems that can be deployed
in order to detect such anamolous behavior.

\begin{table*}
		{\tiny
				\centering
				\begin{tabular}{|c|c|c|c|c|c|c|}
						\hline
						metric& oxycodone& oxycontin&hydrocodone &vicodin &percocet &codeine \\
						\hline
						accuracy            &0.9457     &0.9451& 0.8449 & 0.9658 & 0.9565   & 0.9342\\
						average precision   &0.9621     &0.9582& 0.8930&  0.9749 & 0.9740 & 0.9673\\
						f1-score            &0.9451     &0.9455 & 0.8342& 0.9656 & 0.9551& 0.9443\\
						precision           &0.9573      &0.9414& 0.8949& 0.9692& 0.9837& 0.9842\\
						recall              &0.9337     &0.9500 & 0.7822& 0.9634& 0.9285 & 0.9023\\
						zero-one-loss       &0.0542     &0.0548&  0.1551& 0.0314& 0.0434&  0.0232\\
						\hline
				\end{tabular}
				\caption{This illustrates the classification results for different drugs
				using logistic regression.}
				\label{table:classification}
		}
\end{table*}

\section{Discussion and Conclusion}
\label{sec:conclusion}
In this work, we developed a methodology to isolate tweets which promote and market
illicit online pharmacies.  We also identified and studied the unique markers of such content and the
users who generate it, and demonstrated that these unique markers could help
idenitify the rogue tweets from unseen data.  A simple machine classifier trained on
numerical features was used to predict
rogue tweets on unseen data.  However, the scenario of marketing and promotion for IOPs
are perhaps constantly evolving.  The perpetrators of such messages might adopt newer strategies
of promotion and reaching out to the users.  Hence, there is a need for more sophisticated 
online learning techniques where intelligent systems can automatically learn the nuances
and adapt accordingly. 

\bibliographystyle{plain}
\bibliography{refs}

\end{document}